\documentclass[preprint,prb,showpacs,amsmath,amssymb,superscriptaddress]{revtex4}
\usepackage{graphicx}
\usepackage{dcolumn}
\usepackage{bm}
\usepackage{amsfonts}
\usepackage{amsmath}
\usepackage{amssymb}

\begin{document}
\title{Aging of poled ferroelectric ceramics due to relaxation of random depolarization
fields by space-charge accumulation near grain boundaries}
\author{Yu.A.~Genenko}
\email{yugenen@tgm.tu-darmstadt.de}
\affiliation{Institut f\"ur Materialwissenschaft, Technische Universit\"at Darmstadt,
64287 Darmstadt, Germany}%
\author{J.~Glaum}
\affiliation{Institut f\"ur Materialwissenschaft, Technische Universit\"at Darmstadt,
64287 Darmstadt, Germany}%
\author{O.~Hirsch}
\affiliation{Institut f\"ur Materialwissenschaft, Technische Universit\"at Darmstadt,
64287 Darmstadt, Germany}%
\author{H.~Kungl}
\affiliation{Institut f\"ur Keramik im Maschinenbau, Universit\"at Karlsruhe, 
76131 Karlsruhe, Germany}%
\author{M.J.~Hoffmann}
\affiliation{Institut f\"ur Keramik im Maschinenbau, Universit\"at Karlsruhe, 
76131 Karlsruhe, Germany}%
\author{T.~Granzow}   %
\affiliation{Institut f\"ur Materialwissenschaft, Technische Universit\"at Darmstadt,
64287 Darmstadt, Germany}%
\date{\today}

\begin{abstract}

Migration of charged point defects triggered by the local random depolarization field 
is shown to plausibly explain aging of poled ferroelectric ceramics providing reasonable 
time and acceptor concentration dependences of the emerging internal bias field. The 
theory is based on the evaluation of the energy of the local depolarization field caused 
by mismatch of the polarizations of neighbor grains. The kinetics of charge migration 
assumes presence of mobile oxygen vacancies in the material due to the intentional or 
unintentional acceptor doping. Satisfactory agreement of the theory with experiment on 
the Fe-doped lead zirconate titanate is demonstrated.
\end{abstract}

\maketitle

\section{\label{sec:intro}Introduction}

Many ferroelectric materials exhibit gradual change of parameters with time 
under equilibrium external conditions, especially when acceptor doped. This phenomenon 
called aging is known almost as long as ferroelectrics themselves~\cite{plessner56aging} 
but its mechanism is still highly disputed. Characteristic features of the degradation 
process are the decreasing dielectric constant and the fixed pattern of the polarization 
which hinders repolarization of the material. One of the first ideas of the aging 
mechanism was piling up of the space charge which pins the polarization 
configuration~\cite{Okazaki1962,Okazaki1965,Okazaki1966,Okazaki1969,takahashi70space,%
thomann72stabilization}. Until recently, there was, however, no 
quantitative description of this mechanism which could allow comparison with experiment. 
An alternative and well elaborated concept of aging in acceptor doped ferroelectrics is 
the mechanism of defect dipole reorientation suggested by Arlt et 
al.~\cite{arlt88internal,Lohkamper1990Gauss} and supported in recent 
works~\cite{Warren1996,zhang06aging,Eichel2008dipols,zhang08dipols}. 
This concept allowed reasonable explanation of the time and temperature dependencies of 
the most important parameters of aging, namely, of the emerging internal bias field, 
$E_{ib}$, and of the characteristic aging time, $\tau$. Nevertheless, this theory still 
seems to miss important features of aging concerning its dependence on the doping level. 
The orientation of the defect dipole due to the random walk of an oxygen vacancy about 
an acceptor defect is assumed to be a microscopic process independent of the other defect 
dipoles. Thus, the aging time appears to be independent of the doping level. The 
internal bias field proportional to the sum of independent contributions of the 
individual dipoles is expected in this theory to be proportional to the concentration of 
acceptor defects, $c_0$. Experimentally, however, aging time is distinctly $c_0$
dependent~\cite{arlt88internal,carl78electrical}, and the internal bias field saturates 
with increasing concentration~\cite{carl78electrical,takahashi82} well below 
$c_0\simeq 1 \rm \: mol \%$. For certain dopants the reason of this saturation could be 
the solubility limit in the host crystal as is stated to be the case for Fe-ions in  
$\rm \: PbZr_x Ti_{1-x}O_3$ (PZT)  ceramics~\cite{Weston1969,Kleebe2009}. The saturation 
of the bias field is observed, however, for virtually all acceptor dopants below 
$c_0\simeq 1 \rm \: mol \%$\cite{carl78electrical,takahashi82} and seems to be a 
universal feature of acceptor doped ferroelectrics. 

As was recently shown, an alternative, charge defect migration mechanism can 
quantitatively explain essential features of aging in unpolarized 
ceramics~\cite{lupascu06aging,Genenko-ferro2007,Genenko-ferro2008} as 
well as fatigue under a constant electric field~\cite{Balke2009}. Following the latter 
concepts we advance in this work a model of aging in poled ferroelectric ceramics due to 
the depolarization-field driven charge migration. Theoretical results are then compared
with model experiments on the PZT ceramics with controlled Fe doping.

\section{\label{sec:generalmodel} Model of fully polarized ferroelectric ceramics}

Ferroelectric ceramics are characterized by at least three sorts of randomness: random 
form of grains, their positions and random orientation of the crystal lattice inside 
the grains. We assume the last one to be the most important factor of randomness which 
can capture the main features of aging and fatigue in these systems, therefore only this 
kind of disorder will be considered in the following. 

\begin{figure}[!tbp]
\begin{center}
    \includegraphics[width=8cm]{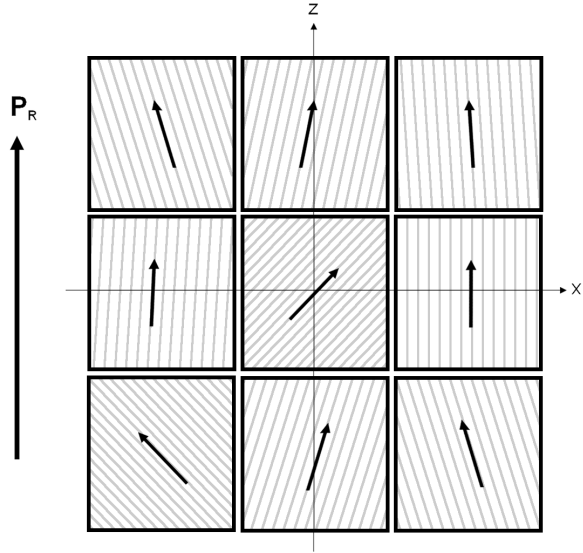}
    \caption{Two-dimensional projection of polarization distribution in a fully polarized
             ferroelectric ceramics. Hatching pattern shows schematically the orientation
             of the crystal lattice.}   
\label{cubic-array}
\end{center}
\end{figure}

We imagine the sample to consist of a regular cubic lattice of equal tightly 
contacted single-crystalline cubic grains of mesoscopic size $R$ much larger than the 
lattice constant of the material. The grain edges are supposed to be aligned along the 
axes of the Cartesian coordinate system $x,y,z$ as is shown in Fig.~\ref{cubic-array}. 

The sample is supposed to be sandwiched between plane electrodes located at $z=\pm L/2$, 
where $L\gg R$, and polarized in a dc electric field substantially higher than the 
coercive field to the maximum possible spontaneous polarization in the $z-$direction. 
After that the voltage at the electrodes is set back to zero so that the remanent 
polarization ${\bf P}_r$ in the $z-$direction remains. This state is considered as 
initial one for the process of aging at a fixed temperature and zero voltage maintained.

Since the crystal structure of the grains is formed at temperatures far above the 
ferroelectric phase transition the crystal axes orientation in different grains is 
supposed to be absolutely arbitrary and independent. In the high electric field the
polarization in every grain takes on the direction of one of the pseudocubic axes most
close to the direction of the applied field (see in Fig.~\ref{cubic-array}). 
Note that the assumption that each grain has a homogeneous polarization is quite strong, 
as in reality grains always consist of a multitude of domains with different polarization 
directions. This is a well-known simplification that is often used in 
literature~\cite{Baerwald1957,Uchida1967}. Thus, the vectors ${\bf P}_s$ of local 
spontaneous polarization have the same magnitude of $P_s$ and are randomly distributed 
within the cone defined by the polar angle $\theta <\theta_{max}$ with respect to the 
$z-$axis where $\theta_{max}=\arcsin{(\sqrt{2/3})}$ is the threshold angle introduced 
by Uchida et al.~\cite{Uchida1967}. Further simplifying the problem we assume that the 
achieved maximum polarization in the field direction remains after setting the applied 
voltage back to zero. For calculation of thermodynamic characteristics one needs a 
procedure of statistical averaging described in the next section.

\subsection{\label{subsec:randomness}Configurational averaging}

Configurational averaging of local angle-dependent quantities over the ensemble of all
possible random configurations, equal to the averaging over the sample volume, may be 
performed using the distribution function of a possible polarization direction in an 
arbitrary grain given by
\begin{align}
\label{single-distribution}
f(\theta,\varphi )&= \frac{3}{2\pi },\,
\,0\leq \theta \leq \pi /4\, \,\,\,\,\,\,\,\, \text{and}   \nonumber\\
&= \frac{6}{\pi^2} \left[ \frac{\pi }{4} - \arccos{(\cot{\theta })} \right],
\pi /4 \leq \theta \leq \theta_{max}  
\end{align} 
for $0\leq \varphi<2\pi$ where $\varphi$ and $\theta$ are the azimuthal and polar angle 
in spherical coordinates associated with the above introduced cartesian coordinates and 
centered in the center of the chosen grain. The distribution function $f(\theta,\varphi)$ 
is derived in Appendix~\ref{sec:Distribution}. 

Let us introduce three-dimensional numeration (integer coordinates) of grains $[n,k,m]$ 
associated with the cartesian coordinates so that the centers of the grains take 
positions $(nR,kR,mR)$. Assuming independence of random angle variables in different 
grains the distribution function for polarization directions in all grains reads
\begin{equation}
\label{multy-distribution}
F(\{\theta_i,\varphi_i\} )= \prod_{n,k,m} 
f(\theta_{n,k,m},\varphi_{n,k,m} )
\end{equation} 
where $\{\theta_i,\varphi_i\}$ denotes the manifold of spherical angles in all grains,
while $\theta_{n,k,m}$ and $\varphi_{n,k,m}$ denote the angles in the grain with numbers
$[n,k,m]$.

Calculating an ensemble average of a quantity $g(\{\theta_i,\varphi_i\})$: 
\begin{equation}
\label{average}
<g>= \prod_{i}\int \sin{(\theta_{i})} d\theta_{i} \int\ d\varphi_{i}\, 
 g(\{\theta_i,\varphi_i\}) 
F(\{\theta_i,\varphi_i\} )
\end{equation} 
one should
take into account that all azimuthal and polar angle variables change in the same ranges
as respective angles in Eq.~(\ref{single-distribution}). Thus, the ensemble average of 
the polarization along the $z-$axis, $<P_z>$, amounts to
\begin{equation}
\label{ave-Pz}
P_r = P_s <\cos{(\theta_{n,k,m}})>=P_s\frac{3\sqrt{2}}{\pi}\arcsin{\sqrt{1/3}},
\end{equation} 
and equals $0.831 P_s$ as in Refs.~\cite{Baerwald1957,Uchida1967}
while the perpendicular polarization component averages to
\begin{equation}
\label{ave-Pxy}
<P_x> = P_s <\cos{(\theta_{n,k,m}})\cos{(\varphi _{n,k,m}})>=0,
\end{equation} 
as well as $<P_y>=0$. The local polarization can be then conveniently decomposed in
a sum of the mean and fluctuation polarizations as
${\bf P}_s = {\bf P}_r + \Delta {\bf P}_s$ where apparently $<\Delta {\bf P}_s>=0$.

\subsection{\label{subsec:mean} Mean values and variances of charges and fields} 

Surface bound charge densities at the faces of a cubic grain with the number $[n,k,m]$ 
located inside the bulk material are constant over the cubic faces and result from 
discontinuities of the respective normal components of the polarization $\Delta{\bf P}_s$
in the neighbor grains, namely, the charge density at the bottom face perpendicular to 
the axis $z$ equals
\begin{equation}
\label{sigma-z}
\sigma^z_{n,k,m}=P_s \left( \cos{(\theta_{n,k,m-1})} - \cos{(\theta_{n,k,m})}\right),
\end{equation} 
the charge density at the left face perpendicular to the  axis $x$ equals
\begin{align}
\label{sigma-x}
\sigma^x_{n,k,m} & =
P_s \left[ \sin{(\theta_{n-1,k,m})}\cos{(\varphi _{n-1,k,m})}\right.\nonumber\\
& \left. - \sin{(\theta_{n,k,m})}\cos{(\varphi _{n,k,m})} \right],
\end{align} 
and the charge density at the left face perpendicular to the  axis $y$ equals
\begin{align}
\label{sigma-y}
\sigma^y_{n,k,m} & =
P_s \left[ \sin{(\theta_{n,k-1,m})}\sin{(\varphi _{n,k-1,m})}\right.\nonumber\\
& \left. - \sin{(\theta_{n,k,m})}\sin{(\varphi_{n,k,m})} \right],
\end{align} 
Configurational averaging of the above charge densities for internal grains with the 
function (\ref{multy-distribution}), equal to the averaging over the sample volume, 
results in vanishing mean values $<\sigma^{x,y,z}_{n,k,m}>=0$, which does not preclude 
the fact that local values (\ref{sigma-z},\ref{sigma-x},\ref{sigma-y}) are finite. 
For the top plane of the sample Eq.~(\ref{sigma-z}) is not valid because there
are no grains above the top grain layer. For that reason the nonzero mean value
$<\sigma^z_{n,k,m}>=P_r$ produced by the mean polarization ${\bf P}_r$ prevails at the 
top plane of the sample, $z=L/2$. Similarly, $<\sigma^z_{n,k,m}>=-P_r$ at the bottom 
plane of the sample, $z=-L/2$.

Typical magnitudes of the charge densities at the internal grain faces 
(\ref{sigma-z},\ref{sigma-x},\ref{sigma-y}) are characterized by the variances of the 
respective charge densities and amount to
\begin{align}
\label{charge-variances}
\bar \sigma^z &= \sqrt{<(\sigma^{z}_{n,k,m}  )^2>} = 0.142\cdot P_s
\nonumber \\
\bar \sigma^{x,y} &=\sqrt{<(\sigma^{x,y}_{n,k,m}  )^2>} = 0.547\cdot P_s
\end{align} 
The substantial difference between $\bar \sigma^z$ and $\bar \sigma^{x,y}$ is explained 
by the fact that the direction of the polarization in the $(x,y)-$plane and, hence, 
variation of its azimuthal angle are not restricted while the polar angle is confined to 
the cone $\theta \leq \theta_{max} $ around the positive $z-$direction. 

Consistently with the splitting of polarization the total electric depolarization field 
can be decomposed as ${\bf E}_d=<{\bf E}_d>+\Delta{\bf E}_d$ where the mean 
depolarization field $<{\bf E}_d>$ is created by the constant charge densities $\pm P_r$ 
at the planes $z=\pm L/2$, resulting from the mean polarization ${\bf P}_r$, and the 
fluctuation field $\Delta{\bf E}_d$ is created by the fluctuation polarization 
$\Delta {\bf P}_s$. In the considered poled state subject to aging, the  field ${\cal E}$ 
created inside the ferroelectric by the external sources (electrodes at $z=\pm L/2$) 
exactly matches the mean depolarization field so that the mean total electric field 
$<{\bf E}>={\cal E}+<{\bf E}_d>$ vanishes together with the voltage at the electrodes, 
$V=0$. This means that the mean surface charge densities at the top and bottom planes of 
the sample are precisely compensated by the opposite charge densities $\sigma =\mp P_r$ 
at the respective electrodes.  

Local depolarization field $\Delta{\bf E}_d$ is connected by a nonlocal linear relation 
with the charge densities at all grain faces. To investigate statistic properties of this 
field we consider in the following exemplary the field in the center of the central grain 
with numbers $[0,0,0]$ which is representative for all internal grains of bulk 
ferroelectric ceramics. This field is given by the tensor relation
\begin{equation}
\label{field}
\Delta E^{\alpha}_d=\frac{-1}{4\pi \epsilon_0 \epsilon_f}\sum_{n,k,m}\sum_{\beta}
T_{n,k,m}^{\alpha \beta}\sigma^{\beta }_{n,k,m}
\end{equation} 
derived in the Appendix~\ref{sec:Field}. Here indices $\alpha ,\beta $ are introduced, 
taking on values $x,y,z$, and matrix elements $T_{n,k,m}^{\alpha \beta}$ are defined by 
the integrals over the cubic faces as is shown in the Appendix~\ref{sec:Field}. The 
dielectric tensor of the ferroelectric medium is assumed to be of virtually cubic 
symmetry and is characterized by the relative permittivity $\epsilon_f$. 

Typical magnitudes of the local electric field components are given by respective 
variances:
\begin{align}
\label{varfield}
&<(\Delta E^{\alpha}_d)^2>=\left(\frac{P_s}{4\pi \epsilon_0 \epsilon_f}\right)^2
\times\\
&\sum_{n,k,m} \sum_{n',k',m'} \sum_{\beta,\beta'}T_{n,k,m}^{\alpha \beta}
 T_{n',k',m'}^{\alpha \beta'}<\sigma^{\beta }_{n,k,m}\sigma^{\beta' }_{n',k',m'}>
\nonumber
\end{align} 
which are calculated in the Appendix~\ref{sec:Field}. The resulting typical field
deviations are
\begin{align}
\label{meanvar}
\sqrt{<(\Delta E^{x,y}_d)^2>}&=3.523 \frac{P_s}
{4\pi \epsilon_0 \epsilon_f}\nonumber\\ 
\sqrt{<(\Delta E^{z}_d)^2>}&=0.898 \frac{P_s}
{4\pi \epsilon_0 \epsilon_f}.
\end{align} 

The preceding analysis demonstrates that local depolarization fields of the typical 
magnitude of $P_s/4\pi \epsilon_0 \epsilon_f$ are present in bulk of the grains  
in the virgin state of poled ferroelectric ceramics. For the PZT materials with 
$P_s\simeq 0.45 \rm \: C/m^2$, and the lattice (high-field) value of the permittivity 
about $\epsilon_f\simeq 600$ this field is about $2\cdot 10^7 V/m$.

\subsection{\label{subsec:Dynamics} Evolution of the electric field in the ceramics
during aging} 

If mobile charge carriers are present in the ceramics they have to be driven by the 
local electric field. Perovskite ferroelectrics are known to be semiconductors with 
predominantly electronic or hole conductivity in reducing or, respectively, oxidizing 
atmosphere~\cite{Okazaki1969,DMSmyth1994,Brennan1995,Raymond1996perovskitechemistry,%
DMSmyth2003,waser91bulk,Molak_PRB2008,Molak_pss2009}, and a noticeable ionic 
contribution to the conductivity under the intermediate conditions. Specifically, PZT 
exhibits domination of ionic conductivity in a wide temperature range at atmospheric 
oxygen pressure~\cite{Raymond1996perovskitechemistry}. In any case, the density of 
electronic carriers in the samples quenched from high temperatures to the room 
temperature is by many orders of the magnitude not sufficient to screen the surface 
bound charges in the ceramics~\cite{Brennan1995,Genenko-ferro2008}. In contrast to this, 
the density of oxygen vacancies, presumably equal to half the concentration of acceptor 
defects for electroneutrality reasons, is rather large since ferroelectrics are usually 
unintentionally acceptor doped~\cite{Brennan1995,Raymond1996perovskitechemistry,DMSmyth2003} 
with small di- and trivalent cations which substitute for Zr$^{4+}$ or Ti$^{4+}$. In 
this work the controlled Fe doping of PZT in the range of $0.1 - 1 \rm \: mol \%$ 
is maintained that results in the oxygen vacancies density of 
$c_0\simeq 10^{19} - 10^{20} \rm \: cm^{-3}$, sufficient for screening of 
spontaneous polarization. An important question concerning the space-charge migration 
mechanism is: which part of the oxygen vacancies introduced by the acceptor doping is 
mobile. Measurements of ionic conductivity in Ca-doped barium 
titanate~\cite{Raymond1996perovskitechemistry,DMSmyth2003} 
and in Ni-doped strontium titanate~\cite{waser91bulk} show a substantial 
increase of the conductivity with increasing doping which suggests that a significant
part of the introduced vacancies is mobile. On the other hand, the ionic conductivity
rises subproportionally with doping which may be the effect of defect 
association~\cite{waser91bulk}. In the following we consider oxygen vacancies, the most 
mobile ionic species in perovskites, as a suitable agent for slow screening process. 
A possible effect of the subproportional increase of the mobile vacancy concentration 
with doping will be considered when adjusting theoretical results to the experimental 
data.  

During the migration of charge carriers they pile up at the charged grain faces to
compensate the source of the electric field and thereby produce themselves the electric 
field ${\bf E}_M$. This process stops when the surface bound charges 
$\sigma^{\alpha}_{n,k,m}$ are outweighed by the emerging space charge. The charge 
redistribution inside the grains may be interpreted as a polarization ${\bf P}_M$, 
superimposed over the local polarization $\Delta {\bf P}_s$. This is indeed a sort of 
migration polarization systematically studied experimentally by Okazaki in 
Refs.~\cite{Okazaki1962,Okazaki1965,Okazaki1966,Okazaki1969}. In our model, however,
this migration is induced by the fluctuation depolarization field 
$\Delta{\bf E}_d$ which was not considered in the mentioned works by Okazaki. 
Typical thickness of space charge zones near the faces about 
$h \simeq \bar \sigma^{\alpha }/qc_0$, with the oxygen vacancy charge equal  
twice the elementary charge $q$ and the vacancy concentration equal to half the
acceptor concentration $c_0$, amounts to $2 - 10$ nm at $c_0 = \rm\: 1 mol\%$. This 
thickness is by two orders of magnitude less than the grain size $R$. This means
that the space charge density $\sigma_M = h\nabla {\bf P}_M$ related to the migration 
polarization may be simply included in the total time dependent surface charge 
$\sigma^{\alpha} _{n,k,m}(t)=\sigma^{\alpha} _{n,k,m}(0)+ \sigma_M(t)$
where initial values at $t=0$ are given by the virgin charge 
densities~(\ref{sigma-z},\ref{sigma-x},\ref{sigma-y}).

The influx of the screening charge to a charged face is driven by the local value of 
the electric field at the face which consists of the field generated by the face 
itself and that induced by the other faces. The latter contribution is continuous 
across the face and does not result in the net change of the surface charge. The former 
component is normal to the face at its both sides and equals 
$\sigma^{\alpha} _{n,k,m}/2\epsilon_0 \epsilon_f$ according to Gauss theorem.   
Thus the charge changes according to equation
\begin{equation}
\label{relaxation}
\partial_t \sigma^{\alpha} _{n,k,m}= -\kappa\sigma^{\alpha} _{n,k,m}/\epsilon_0 \epsilon_f
\end{equation}
where $\kappa=q\mu c_0$ is the conductivity due to oxygen vacancies with the mobility
$\mu $.  
Consequently, the complicated and random spatial distribution of the depolarization field 
in the sample remains unchanged during the charge migration while the magnitude of this 
field scales down coherently at all charged faces of the grains as
$\sigma^{\alpha} _{n,k,m}(0)\exp{(-t/\tau_r )}$ where 
$\tau_r =\varepsilon_0 \varepsilon_f / \kappa$ is the Maxwell-Wagner relaxation time.
The latter seems to be the only characteristic time for aging mechanism through the 
charge migration assuming homogeneous background density of acceptor defects. 

In fact, 
there can be reasons for distribution of this time in a rather wide range. For example, 
if the activation energy of mobility is randomly distributed in some range this can 
result in quasilogarithmic time dependence of material parameters as in the case of 
aging in unpolarized ferroelectrics~\cite{Lohkamper1990Gauss,Genenko-ferro2008}. Another 
reason can be the field dependence of mobility as well as the complicated interplay of 
acceptor charge states and electronic state occupancies because of the band bending by 
the strong electric field near the charged faces of the grains as was suggested in 
photochemical studies of PZT~\cite{Gallardo2008}. The latter phenomena are still beyond 
the scope of our simple model. 

The relaxation time is determined by the concentration and mobility of the  
oxygen vacancies. The former value is controlled by the acceptor concentration and can 
be reduced by the possible defect association as was mentioned before. The latter value 
is still highly disputed. Despite of the wide consensus concerning the migration energy 
barrier for oxygen vacancies about 
$E_a=0.9 - 1.1\rm \: eV$~\cite{Raymond1996perovskitechemistry,DMSmyth2003,waser91bulk,Gottschalk2008}
the reported experimental values of the mobility at temperature $T= 250^{\circ}\rm \: C$
range from $8\cdot 10^{-9} \rm \: cm^2/Vs$ , obtained by the simultaneous thermoelectric 
power and conductivity measurements in 
Refs.~\cite{Raymond1996perovskitechemistry,DMSmyth2003,waser91bulk}, down to 
$2\cdot 10^{-14} \rm \: cm^2/Vs$ established from the diffusion depth profiles for 
oxygen tracers at the same temperature in Ref.~\cite{Gottschalk2008}. Estimations at room 
temperature and the concentration of $c_0= 1 \rm \: mol\%$ result accordingly in a wide 
range of possible aging times from $10^{3} \rm \: s$ 
to $4\cdot 10^{8} \rm \: s$. In the Ref.~\cite{Gottschalk2008}, however, the strongly 
donor doped ($1 - 4 \rm \: mol\%$ of $\rm \: Nb^{5+}$) PZT ceramics were studied in 
contrast to the acceptor doped ceramics of the 
Refs.~\cite{Raymond1996perovskitechemistry,DMSmyth2003,waser91bulk} 
which could have effect on the oxygen vacancy mobility. In the following we will rely on 
the ionic conductivity measurements of the acceptor doped PZT 
ceramics~\cite{Raymond1996perovskitechemistry}.

\subsection{\label{sec:Thermo}Thermodynamic analysis}

Considering the Gibbs free energy for ferroelectrics, Eq.~(\ref{Gibbs-improved}) of 
Appendix~\ref{sec:Th}, for the sample in the uniform external field exactly 
compensating the mean depolarization field, ${\cal E}=-<{\bf E}_d>$,
the contribution of the conductors $\sim \sigma V$ vanishes because of the
zero voltage at the electrodes, $V=0$,  resulting in the initial energy before 
aging starts: 
\begin{equation}
\label{Gibbs-special}
G = F_0\Omega  -\frac{1}{2}\int dV \Delta {\bf P}_s \Delta {\bf E}_d 
\end{equation} 
where $\Omega $ is the sample volume. The latter term here presents the contribution of the 
local fluctuations of the field and polarization to the total energy of the system and 
equals at $t=0$
\begin{equation}
\label{Gibbs-fluct}
\Delta G_{+}(0)=-\frac{1}{2}\Omega \sum_{\alpha }<\Delta P_s^{\alpha }
\Delta E_d^{\alpha }>=0.649  \frac{P_s^2\Omega}{4\pi \epsilon_0 \epsilon_f}
\end{equation}
as is evaluated in the Appendix~\ref{sec:Field}. 
In the course of aging the local field and polarization are modified as
$\Delta {\bf P}_s \rightarrow \Delta {\bf P}_s + {\bf P}_M(t)$
and $\Delta {\bf E}_d\rightarrow \Delta {\bf E}_d + {\bf E}_M(t)$ and
decrease as $\sim \exp{(-t/\tau_r )}$. Thereby the fluctuation contribution to 
the energy becomes
\begin{equation}
\label{Gibbs-aged}
\Delta G_{+}(t)=-\frac{1}{2}\int dV  [\Delta {\bf P}_s+ {\bf P}_M(t)] 
[\Delta {\bf E}_d+ {\bf E}_M(t)]. 
\end{equation}
This energy decreases with time $\sim \exp{(-2t/\tau_r )}$ from the initial positive
value, Eq.~(\ref{Gibbs-fluct}), driving the system to an energy minimum which means 
pinning of the given macroscopic polarization state. The strength of this pinning is 
characterized by the internal bias field which can be evaluated as follows.   

When reversing the external field to the opposite direction after the aging time $t$ the 
local polarization and field change as ${\bf P}_s \rightarrow -{\bf P}_s $ and 
${\bf E}_d\rightarrow -{\bf E}_d$. Thereby the fluctuation contribution to the energy 
changes to
\begin{equation}
\label{Gibbs-reverse}
\Delta G_{-}(t)=-\frac{1}{2}\int dV  [-\Delta {\bf P}_s+ {\bf P}_M(t)] 
[-\Delta {\bf E}_d+ {\bf E}_M(t)]
\end{equation} 
where ${\bf P}_M(t)$ and ${\bf E}_M(t)$ remain unchanged because the charge defects are 
too slow to follow the repolarization immediately. Taking into account that 
${\bf P}_M(t)=-\Delta {\bf P}_s [1-\exp{(-t/\tau_r)}]$ and
${\bf E}_M(t)=-\Delta {\bf E}_d [1-\exp{(-t/\tau_r)}]$
the difference in the energy of the opposite poled states amounts to 
\begin{align}
\label{Gibbs-differ}
\Delta G_{-}(t)-\Delta G_{+}(t)& =\int dV  
[\Delta {\bf P}_s {\bf E}_M(t) + {\bf P}_M(t) \Delta {\bf E}_d]\nonumber\\ 
& =4 \Delta G_{+}(0)[1-\exp{(-t/\tau_r)}]. 
\end{align} 
Considering the increase in energy, which should be overcome by the repolarization, 
one can introduce an effective internal bias field as it was done by Arlt et 
al.~\cite{arlt88internal}
\begin{align}
\label{bias}
E_{ib}^{\parallel }(t)&= \frac{[\Delta G_{-}(t)-\Delta G_{+}(t)]}{P_s\Omega }\nonumber\\  
&=A \frac{P_s}{4\pi \epsilon_0 \epsilon_f}\left[1-\exp{(-t/\tau_r )}\right]
\end{align}
where $A=2.597$ was calculated using Eq.~(\ref{Gibbs-fluct}). The maximum value of this 
field achieved at times $t\gg \tau_r$ is about  $1.7\cdot 10^7 \rm \: V/m$ which is in
agreement with experimental estimations in Ref.~\cite{Okazaki1969} but exceeds by one
order of the magnitude the values reported in Refs.~\cite{carl78electrical,takahashi82}.
Very high theoretical value of the saturated bias field may follow from the basic 
hypothesis of the model on the absence of any correlations between polarizations in 
neighbor grains. That assumption allows appearance of unfavorable local configurations 
with high charge at the grain faces which can in reality be substantially depressed by 
local correlations. This fact will be considered below by adjusting the experimental 
curves. 

When applying to the aged sample the external field in the perpendicular direction as
it is done in some experiments~\cite{zhang06aging} the fluctuation contribution becomes
\begin{equation}
\label{Gibbs-quer}
\Delta G_{\perp }(t)=-\frac{1}{2}\int dV  [\Delta {\bf P}^{\perp }_s+ {\bf P}_M(t)] 
[\Delta {\bf E}^{\perp }_d+ {\bf E}_M(t)]. 
\end{equation} 
The fluctuation polarization in this state, $\Delta {\bf P}^{\perp }_s$, is  
correlated neither with the initial fluctuation polarization $\Delta {\bf P}_s$ nor with 
the field $\Delta {\bf E}_d$, therefore $<\Delta {\bf P}^{\perp }_s {\bf E}_M(t)>=0$ and, 
for the same reason, $<{\bf P}_M(t) \Delta {\bf E}_d^{\perp }>=0$. On the other hand, 
$<\Delta {\bf P}^{\perp }_s \Delta {\bf E}_d^{\perp } >=
<\Delta {\bf P}_s \Delta {\bf E}_d >$ since the system is macroscopically isotropic and
this mean value should be direction independent. Consequently, the excess energy by the 
repolarization in perpendicular direction amounts to 
\begin{align}
\label{differ-quer}
\Delta G_{\perp }(t)-\Delta G_{+}(t)&=\frac{1}{2}\int dV  
[\Delta {\bf P}_s {\bf E}_M(t) + {\bf P}_M(t) \Delta {\bf E}_d]\nonumber\\ 
&=2\Delta G_{+}(0)[1-\exp{(-t/\tau_r)}], 
\end{align} 
and the internal bias field in the perpendicular direction is then equal half that
in the initial polarization direction, $E_{ib}^{\perp}(t)=(1/2) E_{ib}^{\parallel}(t)$.    

We stress that the maximum magnitude of the bias field, Eq.~(\ref{bias}), achieved at 
times $t\gg \tau_r$, is defined by the energy of the fluctuation electrostatic field, 
$\Delta G_{+}(0)$. This energy, Eq.~(\ref{Gibbs-fluct}), is proportional to the volume 
of the sample and, thus, is at least one order of the magnitude higher than the 
corresponding energy in the case of the unpolarized ferroelectrics~\cite{Genenko-ferro2008}, 
which is accumulated near the charged domain faces. If the defect concentration in the 
bulk is so large that $c_0\gg c^{\ast }= \bar \sigma^{\alpha }/qR$ and, hence, 
the thickness of the space-charge zone $h \ll R$, the whole energy of the fluctuation 
electrostatic field is virtually suppressed in the course of aging and, therefore, the 
maximum value of $E_{ib}$ becomes concentration independent. In fact, the grain size $R$ 
itself is doping dependent in the considered material which is typical for ferroelectric 
ceramics~\cite{Okazaki1969} so that it is about $10\rm \: \mu m$ at 
$c_0 = 0.1\rm \: mol\%$ and decreases to $1\: \mu m$ at 
$c_0 = 1\rm \: mol\%$. Thus, $h \ll R$ seems to be always the case. An apparent $c_0$ 
dependence of the measured $E_{ib}$ may arise, however, due to the concentration 
dependence of  the relaxation time $\tau_r$. Thus, at the initial stage of aging, 
$t\ll \tau_r$, the bias field $E_{ib}(t)\sim t/\tau_r\sim c_0$.

\section{\label{sec:Exp}Experiment on PZT}

To study the effect of doping on aging, PZT bulk samples with a zirconia to titania ratio 
of 54/46 were prepared with controlled contents of iron by a mixed oxide route 
\cite{Hammer1998}. Undoped samples were used as well as samples doped with 0.1, 0.5 and 
1 mol\% of Fe. The powders were calcined, pressed into cylindrical bodies and sintered at 
$1050^\circ \rm \:C$ in air for 6 hours. The bodies were cut into disc shaped samples 
with a wire saw (Well 2420, Well Diamantdrahtsaegen GmbH, Mannheim, Germany) and polished 
down to $3\rm \: \mu m$ grid size. After polishing the samples were relaxed in a furnace 
at a maximum temperature of $400^{\circ} \rm \:C$ for 2 hours to relieve mechanical 
stresses that were induced by the cutting and polishing process. Silver electrodes of 
$50 \rm \: nm$ thickness were sputtered on the polished samples and a silver paste (Gwent 
Electronic Materials Ltd., Pontypool, U.K.,) was fired on afterwards at 
$400^{\circ} \rm \:C$.

For the measurements it was necessary to start with fully unaged samples to ensure a 
statistical distribution of all defects and free charges. If the samples are cooled 
slowly through the Curie temperature they age very fast and one can only obtain highly 
pinched hysteresis loops~\cite{MorozovThesis}. Therefore the samples were placed short 
circuited in a furnace at  $450^{\circ} \rm \:C$ for 10 minutes before the first 
measurement. This leads to the necessary redistribution of the defects within the 
sample. After that they were quenched in pure water down to room temperature. By this 
the statistical charge configuration is "frozen" and the samples can easily be poled and
measured in the unaged state.

The polarization measurements were done in a Sawyer-Tower circuit with a 
$4.78 \rm \: \mu F$ test capacitance. The samples were placed between two soft 
springs in a silicon oil bath and the bipolar polarization hysteresis loops were measured 
at a frequency $f=55 \rm \: mHz$ and a maximum field amplitude $E=\pm 2 \rm \:kV/mm$. 
The data were recorded by an oscilloscope.  For each measurement step two bipolar cycles 
were applied. The first two cycles lead to the poling of the sample. After poling and 
measuring the samples were short circuited. Until the next measurement they were kept 
at room temperature in a sample box. Data points were taken after various time steps 
between 40 s and $10^6$ s.

The measured P-E loops were evaluated with respect to the internal bias field $E_{ib}$, 
to record the shift of the polarization curve along the field axis depending on the 
aging time \cite{arlt88internal, carl78electrical}. This field was calculated as follows:
\begin{equation}
\label{Eib_calc}
E_{ib}=-(E_c^- + E_c^+)/2
\end{equation}
where $E_c^-$ is the coercive field when a negative external field is applied and $E_c^+$ 
is the coercive field on the positive field side.

\begin{figure}[!tbp]
\begin{center}
    \includegraphics[width=8cm]{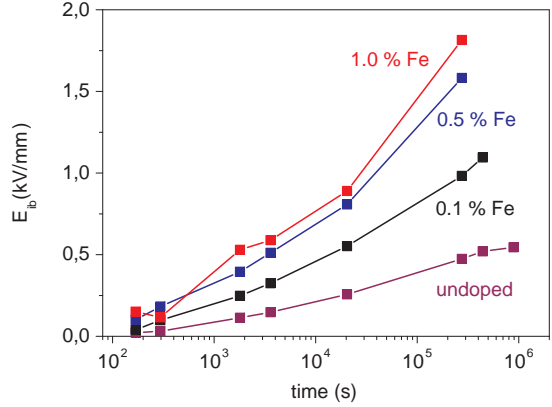}
    \caption{(Color online) Internal bias field as a function of time for PZT samples 
			  with different Fe contents.}   
\label{EIB}
\end{center}
\end{figure}

The internal bias field is presented as a function of time for different Fe contents
in Fig.~\ref{EIB}. 
Two common features of these curves are noticeable: they are virtually quasilinear on the 
logarithmic scale and all of them but one exhibit increasing slope. The first property 
is characteristic of aging and can be explained assuming wide distribution of relaxation 
times in the range $\tau_{min} < \tau _r < \tau_{max}$~\cite{Lohkamper1990Gauss}, the 
second feature means then that the time span of measurements spreads well above 
$\tau_{min}$ but is still well below  $\tau_{max}$. Wide distribution of times $\tau_r$ 
may follow from a random distribution of activation energies 
$E_a$~\cite{Lohkamper1990Gauss,Genenko-ferro2008}
for which there are at least two physical reasons: various positions of the atomic cells 
with respect to the grain boundary, which can affect $E_a$, and the mixture of regions 
with rhombohedral and tetragonal symmetry typical of the systems near the morphotropic 
phase boundary like the compound $\rm \: PbZr_{0.54} Ti_{0.46}O_3$ studied here.
We assume in the following that the activation energy is a random variable with a 
Gaussian distribution $g(E_a)\sim \exp{[-(E_a-\bar E_a)^2/2 s^2}]$ of the width 
$s\simeq 0.22 \rm \: eV$ around the mean value $E_a\simeq 1.12 \rm \: eV$. 
 The internal bias field is then averaged over this 
distribution as
\begin{equation}
\label{Eib-ave}
\bar E_{ib}(t)=\int_0^{\infty } dE_a g(E_a) E_{ib}(E_a,t).
\end{equation} 
where the bias field $E_{ib}(E_a,t)$ is given by Eq.(\ref{bias}) with the relaxation 
time $\tau_r =\varepsilon_0 \varepsilon_f /q\mu c_0$ defined by the 
mobility~\cite{waser91bulk} $\mu\sim T^{-1}\exp{(-E_a/kT)}$ with absolute temperature 
$T$ and the Boltzmann constant $k$.

\begin{figure}[!tbp]
\begin{center}
    \includegraphics[width=8cm]{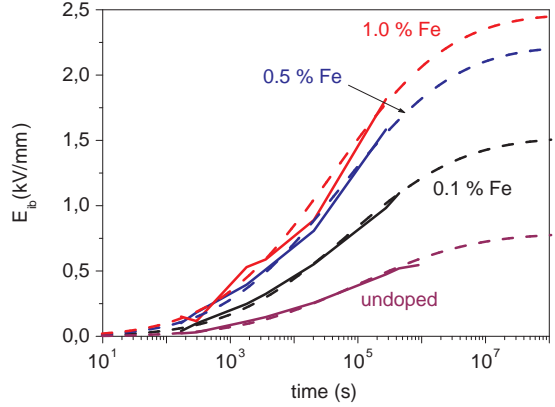}
    \caption{(Color online) Fitting of the experimental time dependencies of the internal 
              bias field of Fig.~\ref{EIB} (solid lines) with theoretical curves of 
             Eq.~(\ref{Eib-ave}) (dashed lines) is performed using the following field 
             reduction factors and the effective concentrations of the mobile vacancies 
             (see text): $\alpha =0.141$ and $c_{eff}=0.15 \rm\: mol\%$ for 
             $c_0=1.0 \rm\: mol\%$, $\alpha =0.127$ and $c_{eff}=0.12 \rm\: mol\%$ for 
             $c_0=0.5 \rm\: mol\%$, $\alpha =0.087$ and $c_{eff}=0.09 \rm\: mol\%$ for 
             $c_0=0.1 \rm\: mol\%$, and $\alpha =0.045$ and $c_{eff}=0.06 \rm\: mol\%$ for 
             the nominally undoped PZT.}  
\label{GBfit}
\end{center}
\end{figure}

By adjustment of the theoretical curves, Eq.~(\ref{Eib-ave}), to the corresponding 
experimental ones for different doping concentrations two facts should be taken into
account. Firstly, the above-mentioned grain size dependence on $c_0$ may have an
effect on the maximum value of $E_{ib}$. Indeed, it is known that larger grains in 
PZT often contain faults such as subgrains in their crystal 
structure~\cite{FarooqSEM2008}. These faults create regions with different polarization 
direction, but the correlation between these regions is higher than between neighboring 
independent small grains. This means higher energy of the fluctuation depolarization 
fields and, accordingly, higher $E_{ib}$ for smaller grains, i.e. for 
higher concentration $c_0$. Secondly, accounting for the subproportional 
increase of the mobile vacancy concentration~\cite{waser91bulk} with increasing 
$c_0$ and for the presumed solubility limit~\cite{Weston1969,Kleebe2009} for Fe in PZT 
at $0.5 - 0.8 \rm \: mol\%$ one should rather use an effective concentration of the 
mobile vacancies $c_{eff}<c_0$ in the formula~(\ref{Eib-ave}). For those reasons, the 
reduction factor for the amplitude of the internal bias field, $\alpha <1$, and $c_{eff}$ 
are used as fitting parameters when adjusting experimental results of Fig.~\ref{EIB} with 
Eq.~(\ref{Eib-ave}). As is seen in Fig.~\ref{GBfit}, satisfactory agreement is achieved 
by $\alpha \simeq 0.05 - 0.1$ and effective concentrations 
$c_{eff}\simeq 0.15 \rm\: mol\%$ for $c_{0}=1.0 \rm\: mol\%$, 
$c_{eff}\simeq 0.12 \rm\: mol\%$ for $c_{0}=0.5 \rm\: mol\%$, 
$c_{eff}\simeq 0.09 \rm\: mol\%$ for $c_{0}=0.1 \rm\: mol\%$ and 
$c_{eff}\simeq 0.06 \rm\: mol\%$ for the nominally undoped material. 
Finally, the presented comparison of the charge migration model with experiments 
demonstrates that this mechanism may provide the observed magnitudes of the internal bias 
field and exhibit a reasonable agreement in the characteristic time and doping dependences.

\section{\label{sec:conclusions}Conclusions}

Even in the state with maximum spontaneous polarization of the ferroelectric ceramics
the polarizations of individual grains cannot perfectly match each other all over the
material because of random and independent orientation of the crystal lattice in 
different grains. That is why local bound charges at the grain surfaces and local 
depolarization field in the bulk are present even if the mean depolarization field is
exactly compensated in the electroded sample by zero voltage applied. We tried to capture
this important feature of ferroelectric ceramics in the model of perfectly ordered cubic 
grains with random crystal lattice orientation in them. The frozen orientational disorder 
of the vectors of polarization in the grains can relax to the ordered state if mobile 
charges are present, which is the case in the acceptor doped (also unintentionally)
ferroelectrics. Driven by the local depolarization field they can partly or 
completely compensate the fluctuation bound charge densities at the grain faces by 
building the thin space-charge zones near these faces. This is equivalent to effective
alignment of the local polarizations in the direction of the total polarization
accompanied by suppression of the fluctuation depolarization field. The described process 
leads system to the energy minimum where it resists its repolarization to another 
direction, i.e. exhibits aging of the poled ferroelectrics. Quantitative analysis of the 
relaxation kinetics in our model and its comparison to the model experiments with 
controlled acceptor contents show reasonable agreement in the magnitude of the internal
bias field and its dependence on time and doping. By adjusting the experimental results
with theoretical dependencies the activation energy of the vacancies was assumed smeared 
around the mean value of $1.1 \rm\: eV$, the concentration of the mobile vacancies 
$c_{eff}$ was taken reduced with respect to the acceptor concentration $c_0$ and the 
depression of the internal bias field by the possible correlation between the neighbor 
grains was taken into account.

The suggested model assumes charge migration in the bulk of the grains which means, in 
fact, drift over the distance $h$, the thickness of the space charge zone. Another 
screening scenario may occur if the vacancies in the intergranular space are relevant to 
aging~\cite{Okazaki1969}. The vacancy diffusion along the grain boundaries is known to be 
two orders of the 
magnitude faster than the bulk diffusion~\cite{Gottschalk2008} but the characteristic 
distance the charged defects should cover in this mechanism is given by the grain size 
$R\gg h$. The significance of the grain boundary contribution to screening seems 
therefore to increase for systems with smaller grains, i.e. by higher acceptor 
concentrations. This mechanism depends, however, on many still questionable factors,
in the first instance, on the concentration of acceptors and oxygen vacancies in grain 
boundaries~\cite{Weston1969,Kleebe2009} and their charge state~\cite{Gallardo2008}.  

Though the fluctuation depolarization field is present and causes charge migration in 
the whole volume of the polarized ceramics the charge defects pile up at the grain 
boundaries making the considered effect a kind of surface phenomenon. This mechanism
does not preclude, however, other possible bulk mechanisms which can contribute to
the aging of ferroelectric materials as, for example, the aforementioned defect dipole
mechanism. Formation of such Fe-O and Cu-O dipoles in the bulk lead titanate could be 
detected by means of electron paramagnetic resonance~\cite{Eichel2008dipols}. It would 
be also interesting to prove by spectroscopic methods the presence of deficit or excess 
free vacancies and, respectively, Fe ions in space charge zones near the grain 
boundaries. This would be, however, not easy to detect because the volume share of
these zones is about $h/R\sim 10^{-3} - 10^{-2}$.

\begin{acknowledgments}
Useful discussions with Karsten Albe, R\"udiger Eichel and J\"urgen R\"odel are 
gratefully acknowledged. This work was supported by the Deutsche Forschungsgemeinschaft 
through the Collaborative Research Center 595.
\end{acknowledgments}

\appendix

\section{\label{sec:Distribution} Random distribution of polarizations in 
grains}

The distribution function $f(\theta, \varphi)$ can be derived following Uchida and
Ikeda~\cite{Uchida1967}. They considered a domain polarized in the positive 
$z-$direction and found a random distribution of the electric field, of the
magnitude much higher than the coercive field, which is compatible with this direction 
of the polarization. Due to four-fold symmetry of the problem it is sufficient to
consider the azimuthal region $-\pi/4 < \varphi <\pi/4$. A possible $90^{\circ}$-rotation 
of polarization from the $z-$ to $x-$axis occurs when the projection of a unit field 
vector on the $z-$axis, $\cos{\theta }$, becomes equal to the projection on the
$x-$axis, $\cos{\theta' }=\sin{\theta }\cos{\varphi }$. This determines the boundary 
of the area on the unit sphere $\theta (\varphi )  = \arctan{(1/\cos{\varphi})}$.
Thus the polar angle at this boundary changes between $\theta=\pi/4$ at $\varphi=0$
and $\theta_{max}=\arctan{(\sqrt{2})}$ at $\varphi =\pi/4 $.Within this area, the 
probability of different field directions which are compatible with the polarization 
along the $z-$axis is uniformly distributed with distribution function 
$p(\theta,\varphi)=3/2\pi $. Averaging of the field projection on the $z-$direction 
results in $<\cos{\theta }>=0.831$, in accordance with Eq.~(\ref{ave-Pz}). 

For the aims of this paper, we need a distribution function $f(\theta,\varphi)$ of the 
possible polarization orientations compatible with the strong electric field applied in 
the positive $z-$axis direction. In contrast to the function $p(\theta,\varphi)$ this
distribution is obviously not $\varphi-$dependent since there is no special direction
in the $x-y$ plane. We assume here that the crystal lattice orientation is 
completely decoupled from the form and orientation of grains. The probability for
the polarization to occur in the solid angle near the polar angle $\theta $ is given
by relation
\begin{equation}
\label{solid}
2\pi \sin{\theta }\, d\theta\, f(\theta,\varphi) = 
8\int_{\varphi_0 }^{\pi /4} d\varphi\, \sin{\theta }\, d\theta \, p(\theta,\varphi).
\end{equation} 
So long as the polar angle $\theta $ is less than $\pi/4 $ the integration over $\varphi$
in the above integral goes from $\varphi_0= 0$ to $\pi /4$ which results in the constant 
value of $f(\theta,\varphi)$. When $\pi /4 < \theta <\theta_{max} $ the integration goes
from $\varphi_0=\arccos{(\cot{\theta })}$ to $\pi /4$ which results in the function 
shown in Eq.~(\ref{single-distribution}).

\section{\label{sec:Field}Calculation of the local depolarization field
and its moments}

The electric field~(\ref{field}) is calculated by direct summation of contributions
from all grain faces charged with densities~(\ref{sigma-z},\ref{sigma-x},\ref{sigma-y}).  
Let us introduce notations $n_{\pm}=n\pm 1/2$, $k_{\pm}=k\pm 1/2$, and 
$m_{\pm}=m\pm 1/2$. Then the matrix elements $T_{n,k,m}^{\alpha \beta}$ in the 
formula~(\ref{field}) are defined by the integrals over the cubic faces in dimensionless 
variables $X=x/R,Y=y/R$ and $Z=z/R$:
\begin{align}
\label{T}
T^{xx}_{n,k,m}&= \int_{k_{-}}^{{k_{+}}}dY \int_{m_{-}}^{{m_{+}}}dZ 
\frac{n_{-}}{ (n_{-}^2+Y^2+Z^2)^{3/2} }\\
T^{xy}_{n,k,m}&= \int_{n_{-}}^{{n_{+}}}dX \int_{m_{-}}^{{m_{+}}}dZ 
\frac{X}{ (X^2+k_{-}^2+Z^2)^{3/2} }\\
T^{xz}_{n,k,m}&= \int_{n_{-}}^{{n_{+}}}dX \int_{k_{-}}^{{k_{+}}}dY 
\frac{X}{ ( X^2+Y^2+m_{-}^2)^{3/2} }\\
T^{yx}_{n,k,m}&=\int_{k_{-}}^{{k_{+}}}dY \int_{m_{-}}^{{m_{+}}}dZ 
\frac{Y}{ (n_{-}^2+Y^2+Z^2)^{3/2} }\\
T^{yy}_{n,k,m}&=\int_{n_{-}}^{{n_{+}}}dX \int_{m_{-}}^{{m_{+}}}dZ 
\frac{k_{-}}{ (X^2+k_{-}^2+Z^2)^{3/2} }\\
T^{yz}_{n,k,m}&=\int_{n_{-}}^{{n_{+}}}dX \int_{k_{-}}^{{k_{+}}}dY 
\frac{Y}{ ( X^2+Y^2+m_{-}^2)^{3/2} }\\
T^{zx}_{n,k,m}&=  \int_{k_{-}}^{{k_{+}}}dY \int_{m_{-}}^{{m_{+}}}dZ
\frac{Z}{ (n_{-}^2+Y^2+Z^2)^{3/2} }\\
T^{zy}_{n,k,m}&= \int_{n_{-}}^{{n_{+}}}dX \int_{m_{-}}^{{m_{+}}}dZ 
\frac{Z}{ (X^2+k_{-}^2+Z^2)^{3/2} }\\
T^{zz}_{n,k,m}&= \int_{n_{-}}^{{n_{+}}}dX \int_{k_{-}}^{{k_{+}}}dY 
\frac{m_{-}}{ ( X^2+Y^2+m_{-}^2)^{3/2} }
\end{align}
After integration one obtains
\begin{align}
\label{Txx}
T^{xx}_{n,k,m}&=\gamma (m_{+},k_{+},n_{-}) - \gamma (m_{-},k_{+},n_{-})\nonumber\\
&-\gamma (m_{+},k_{-},n_{-}) + \gamma (m_{-},k_{-},n_{-}), 
\end{align}
where we introduced the function 
\begin{equation}
\label{function1}
\gamma (n,k,m) = \arctan{\left(\frac{nk}{m\sqrt{n^2+k^2+m^2}}\right)}, 
\end{equation}
and 
\begin{align}
\label{Txy}
T^{xy}_{n,k,m}&=\ln{ \left[ \frac{\left(m_{+}+\sqrt{n_{-}^2+k_{-}^2+m_{+}^2}\right)}
{\left(m_{+}+\sqrt{n_{+}^2+k_{-}^2+m_{+}^2}\right)} \right] }\\
&+ \ln{ \left[ \frac{ \left(m_{-}+\sqrt{n_{+}^2+k_{-}^2+m_{-}^2}\right) }
{ \left(m_{-}+\sqrt{n_{-}^2+k_{-}^2+m_{-}^2}\right) } \right] }.
\end{align}

For symmetry reasons the other components of the tensor $\hat T$ can be expressed as
\begin{align}
\label{T-rest}
T^{yy}_{n,k,m}&= T^{xx}_{k,n,m},& T^{zz}_{n,k,m}= T^{xx}_{m,k,n},\\
T^{xz}_{n,k,m}&= T^{xy}_{n,m,k},& T^{yx}_{n,k,m}=T^{xy}_{k,n,m} ,\\
T^{yz}_{n,k,m}&=T^{xy}_{k,m,n} ,& T^{zx}_{n,k,m}= T^{xy}_{m,n,k},\\
T^{zy}_{n,k,m}&= T^{xy}_{m,k,n}.
\end{align} 

For calculation of the field component variances using Eq.~(\ref{varfield}) the 
correlation functions of the charge densities have to be obtained first. They appear to 
be diagonal in Cartesian indices $\beta ,\beta' $ and involve only next neighbor indices 
$n,k,m$:
\begin{align}
\label{corr}
<\sigma^{x }_{n,k,m}\sigma^{x }_{n',k',m'}> &=P_s^2 a_3\delta_{k,k'} \delta_{m,m'}\\ 
&\times (2\delta_{n,n'}-\delta_{n,n'-1}-\delta_{n,n'+1})\nonumber\\
<\sigma^{y }_{n,k,m}\sigma^{y }_{n',k',m'}> &=P_s^2 a_3\delta_{n,n'} \delta_{m,m'}
\nonumber\\ 
&\times (2\delta_{k,k'}-\delta_{k,k'-1}-\delta_{k,k'+1})\nonumber\\
<\sigma^{z }_{n,k,m}\sigma^{z }_{n',k',m'}> &=P_s^2 (a_1-a_2)\delta_{n,n'} \delta_{k,k'}
\nonumber\\ 
&\times (2\delta_{m,m'}-\delta_{m,m'-1}-\delta_{m,m'+1})\nonumber 
\end{align}
with constants
\begin{align}
\label{angles}
a_1&=<\cos{(\theta_{n,k,m}})^2>=
\frac{1}{3}+\frac{2}{\pi \sqrt{3}}=
0.701\nonumber\\ 
a_2&=<\cos{(\theta_{n,k,m}})>^2=
\frac{18}{\pi^2}\arcsin^2{(\sqrt{1/3})}=
0.691\nonumber\\ 
a_3&=<\sin{(\theta_{n,k,m})}^2\sin{(\varphi_{n,k,m})}^2>=
\frac{\pi -\sqrt{3}}{3\pi} \nonumber\\ 
&= 0.15.
\end{align} 
This is followed by expressions for the field component variances
\begin{align}
\label{varfield-x}
&<(\Delta E^{\alpha }_d)^2>=\left(\frac{P_s}{4\pi \epsilon_0 \epsilon_f}\right)^2
\\ 
&\times\sum_{n,k,m} \left[ a_3 T_{n,k,m}^{\alpha x} 
\left( 2T_{n,k,m}^{\alpha x}-T_{n-1,k,m}^{\alpha x}-
T_{n+1,k,m}^{\alpha x} \right)\right. \nonumber\\ 
&+a_3 T_{n,k,m}^{\alpha y}
\left( 2T_{n,k,m}^{\alpha y}-T_{n,k-1,m}^{\alpha y}-T_{n,k+1,m}^{\alpha y} 
\right)\nonumber\\
&+(a_1-a_2) T_{n,k,m}^{\alpha z}
\left(2T_{n,k,m}^{\alpha z}-T_{n,k,m-1}^{\alpha z}-T_{n,k,m+1}^{\alpha z}\right]
\nonumber
\end{align}
containing well converging series. The sums in the last formula are evaluated 
numerically resulting in formulas~(\ref{meanvar}).
 
The sum of correlation functions between the polarization and field arising in the Gibbs 
energy~(\ref{Gibbs-fluct}) is position independent and can be evaluated, for example, 
for the central grain with $n=k=m=0$ using Eq.~(\ref{field}): 
\begin{align}
\label{corr-PE}
&\Delta G_{+}(0)=-\frac{P_s^2\Omega }{8\pi\epsilon_0 \epsilon_f }
\left[a_3 (T_{1,0,0}^{xx}-T_{0,0,0}^{xx})\right.\\ 
&\left. + a_3 (T_{0,1,0}^{yy}-T_{0,0,0}^{yy})
+(a_1-a_2) (T_{0,0,1}^{zz}-T_{0,0,0}^{zz})
\right]\nonumber
\end{align} 
resulting in Eq.~(\ref{Gibbs-fluct}). \\

\section{\label{sec:Th}Thermodynamic relations for a ferroelectric sample in an 
external electric field}

An expression for the Gibbs free energy of a ferroelectric body subject to an external 
electric field is given by~\cite{LandauElectrodynamicsContinuum} 
\begin{align}
\label{Gibbs-general}
G &= \int dV \left( F_0 - \frac{1}{2}\epsilon_0 {\bf E}(\hat\epsilon_r {\bf E}) 
- {\bf E P_s} \right)
\nonumber\\
&=\int dV \left( F_0 - \frac{1}{2}{\bf E D} - \frac{1}{2}{\bf E P_s} \right) 
\end{align}
where ${\bf E}$ and ${\bf D}={\bf P}_s + \epsilon_0 \hat\epsilon_r {\bf E}$ are the 
total electrical field and the electrical displacement, respectively, ${\bf P}_s$ 
denotes the local spontaneous polarization, and $\hat\epsilon_r$ the relative 
permittivity tensor. The energy $F_0$ depends on temperature and the material density 
only, while piezoelectric contributions are not considered. The integration in this
expression is over the whole space excluding the volume of the conductors creating
the external field.

Using the potential presentation of the electric field ${\bf E}=-\nabla  \phi $
and the fact that $\nabla {\bf D}=0$ due to the absence of the external 
charges one can transform ${\bf E D} = -\nabla (\phi {\bf D})$ and then convert the 
integration of the respective term in Eqs.~(\ref{Gibbs-general}) to the integral
over the surfaces of the conductors creating the external field. Since these
surfaces are at constant potentials $\phi_k $ and the surface charge densities
are given by the normal component of the electric displacement, $D_n $, the 
Gibbs free energy can be transformed to   
\begin{equation}
\label{Gibbs-improved}
G = {\cal F}_0 -  \frac{1}{2}\sum_k Q_k \phi_k - 
\frac{1}{2} \int dV {\bf E P}_s.  
\end{equation} 
where $Q_k$ are the total charges at the conductors numerated by the index $k$, and 
${\cal F}_0$ depends only on temperature and the material density. 
In contrast to Eq.~(\ref{Gibbs-general}) the integration in the last equation is over 
the volume of the dielectric body only since outside of the body  the polarization 
vanishes. In equilibrium, the Gibbs free energy is minimum with respect to parameters 
involved~\cite{LandauElectrodynamicsContinuum} when the potentials $\phi_k $ at the 
conductors are kept constant.

\bibliographystyle{plain}
\bibliography{apssamp}

\end{document}